\documentclass[aps,pre,reprint,superscriptaddress]{revtex4-1}
\usepackage{amsmath}
\usepackage{amsfonts}
\usepackage{verbatim}
\usepackage{amsthm}
\usepackage{amssymb}
\usepackage{graphicx}
\usepackage{array}
\usepackage{enumitem}
\usepackage{etoolbox}

\newcommand{\fig}{Fig.\ }
\newcommand{\eq}{Eq.\ }
\newcommand{\tab}{Table \ }

\begin{document}

\title{High fraction of silent recombination in a finite population two-locus neutral birth-death-mutation model}

\author{A. B. Melka}
\affiliation{Department of Mathematics, Bar-Ilan University, Ramat Gan 52900, Israel}

\author{Y. Louzoun}
\email[Corresponding author: ]{louzouy@math.biu.ac.il}
\affiliation{Department of Mathematics, Bar-Ilan University, Ramat Gan 52900, Israel}
\affiliation{Gonda Brain Research Center, Bar-Ilan University, Ramat Gan 52900, Israel}

\date{\today}

\MessageBreak

\begin{abstract}
A precise estimate of allele and haplotype polymorphism is of great interest in theoretical population genetics, but also has practical applications, such as bone marrow registries management. Allele polymorphism is driven mainly by point mutations, while haplotype polymorphism is also affected by recombination. Current estimates treat recombination as mutations in an infinite site model. We here show that even in the simple case of two loci in a haploid individual, for a finite population, most recombination events produce existing haplotypes, and as such are silent. Silent recombination considerably reduces the total number of haplotypes expected from the infinite site model for populations that are not much larger than one over the mutation rate. Moreover, in contrast with mutations, the number of haplotypes does not grow linearly with the population size. We hence propose a more accurate estimate of the total number of haplotypes that takes into account silent recombination. We study large-scale Human Leukocyte Antigen (HLA) haplotype frequencies from human populations to show that the current estimated recombination rate in the HLA region is underestimated. 
\end{abstract}


\maketitle

\section{Introduction}

Multiple genetics models relate allele frequencies to their populations' dynamics \cite{moran1958random, kimura1964number, wright1969evolution, ewens1972sampling, watterson1975number}, typically including processes such as mutations, genetic drift, selection, or migration between sub-populations \cite{kimura1962probability, krone1997ancestral, kingman1982genealogy, constable2015stationary}. For the haplotype frequencies, another essential process to consider is recombination \cite{meselson1975general, barton1995general, hudson1983properties}. During cell division in sexual reproduction, crossovers can occur between the maternal and paternal homologous chromosomes and result in the exchange of genetic material \cite{neale2006clarifying, san2008mechanism, heyer2010regulation}. Therefore, offspring may have different combinations of genes than either of their parents on the same chromosome, leading to the creation of new haplotypes and increasing genetic variability. Determining the recombination rate (i.e. the probability for those crossovers to occur) is crucial in evolutionary biology and medical population genetics \cite{aarnink2014deleterious, coop2007evolutionary}. It also has important implications for transplant donors registry management \cite{yunis1971three, lobkovsky2019multiplicative}. 

Two approaches have been proposed to build recombination maps and estimate the recombination rate. The first one, referred to as the direct approach, is strictly experimental and consists of sperm genotyping \cite{cullen2002high}. The second approach is an indirect method that uses genetic linkage (co-inheritance of markers in families) to produce recombination maps for chromosome segments \cite{mcvean2004fine}. These maps describe the distance between genes, or markers, as a function of their probability to recombine. If two genes are on two different chromosomes or very distant, they are considered uncorrelated and the distribution of the haplotypes reflects this independence. On the other hand, two adjacent genes will have a high probability of being inherited together. This probability is a direct function of the recombination rate between those genes and the distance between them \cite{morgan1898developmental}. These segments can then be linked to provide estimates of recombination frequencies for specific chromosomes \cite{begovich1992polymorphism}, typically using maximum likelihood estimation (MLE) \cite{hudson1987estimating, fu1993maximum} based on a coalescent tree model \cite{hudson1988coalescent, mcvean2002coalescent, fearnhead2001estimating, stumpf2003estimating}. The coalescent model goes backward to estimate the time for two individuals to reach their most recent common ancestor. However, this model does not take into consideration silent recombination producing twice the same offspring. The recombination rate computed in the direct approach is per cell division, whereas the rate computed in the indirect approach is per generation. 

In a single gene, fixed population, neutral model, the mutation rate has been related to the number of alleles through $\theta = 4 N_{e} \mu$ (where $\theta$ is the overall number of mutations for the population, $N_{e}$ is the effective population as defined by Kimura \cite{kimura1964number}, and $\mu$ is the individual mutation rate per generation). This estimator was first derived by Watterson to describe mutations \cite{watterson1975number}. It is based on an infinite site model (i.e. each mutation creates a new allele). It has often been assumed that recombination behaves like mutations and the same concept was extended to multiple estimates of the recombination rate, where the number of alleles was simply replaced by the number of haplotypes with $\rho = 4 N_{e} r$ (where $\rho$ is the overall number of recombination for the population and $r$ is the individual recombination rate per generation) \cite{hudson1988coalescent, mcvean2002coalescent, fearnhead2001estimating}. The usage of this estimator is limited by the need to determine the effective population $N_{e}$ \cite{tenesa2007recent, leberg2005genetic}. As such, one needs to simultaneously estimate $N_{e}$ and $\mu$ (or $r$). Therefore, studies often either display the ratio between recombination and mutations \cite{kuhner2000maximum} or simply compute $\rho$ and $\theta$ instead of $\mu$ and $r$, separately or jointly \cite{mcvean2002coalescent, fearnhead2001estimating}. Other studies use samples for which the origin of the population is known \cite{blancher2012use}. Note that the number of alleles actually differs from Watterson's estimator. Indeed, multiple corrections were proposed \cite{fu1993maximum, ramirez2009correcting, felsenstein2006accuracy}. The most significant limitation of this estimator is the assumed equivalence between recombination and mutations. As mentioned above, unlike mutations, recombination is drawn from a finite existing pool of alleles. Specifically, given the fat tail of the type size distribution, there is a non-zero probability of reproducing the same combination of haplotypes (silent recombination). As such, the recombination rate expected from Watterson's formula is largely underestimated. We show that for average size populations, the fraction of silent recombination is close to 1. 

We here use a statistical model on the observed alleles and haplotypes distributions and infer the mutation and recombination rates. We use a birth and death process \cite{shem2017solution} rather than a coalescent tree since it allows for the inclusion of silent recombination. We identify two regimes depending on the size of the effective population. For $N_{e} \gg 1 / \mu$, the pool of alleles is large, and the typical type size for each allele is small so that almost all recombination events create a new haplotype and the infinite site assumption holds for recombination. However, for $N_{e} \leq 1 / \mu$, the number of potential recombination creating new haplotypes is limited by an upper bound, induced by the high probability of sampling very frequent alleles. In this regime, almost all recombination events are silent. The fraction of silent recombination grows with the recombination rate. Moreover, the population size where the transition from one regime to the other occurs increases with the recombination rate. Using this insight, we compute the number of different two-locus haplotypes in a population (the haplotype polymorphism), using a revised relationship between the number of alleles and haplotypes and the mutation and recombination rates. As an application, we analyze the distributions of alleles and haplotypes in the HLA locus for human populations and show that the recombination rate is underestimated.

\section{Single Locus and two-locus models} 

As a preliminary step, we focus on a single locus and estimate the number of alleles with respect to the mutation rate. We follow \cite{shem2017solution} and assume a neutral infinite site Moran model with equal birth and death rates (so that the total population is maintained fairly constant) that can be arbitrarily set equal to 1 (up to a time scaling). We define $\mu$ as the per generation mutation rate and the probability for an allele to have a population of size $k$ is given by Fisher log-series \cite{fisher1943relation} (see the Appendix for a simplified derivation). Accordingly, the expected total number of different alleles (richness or first moment of the distribution) is given by $m_{0} = - N \mu \ln{\mu}$.

\begin{figure}[ht]
\begin{center}
\includegraphics[width=0.48\textwidth]{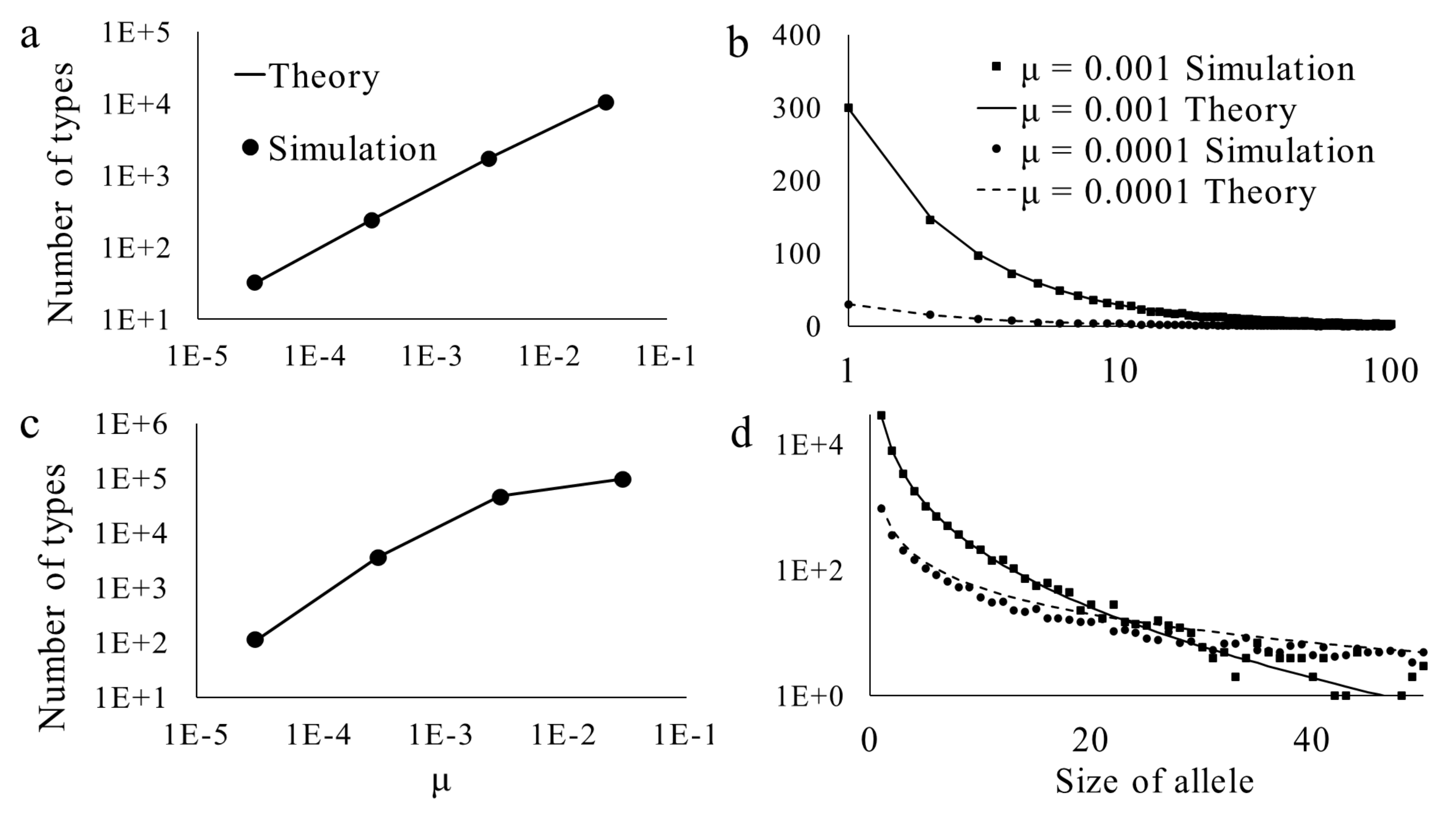}
\caption{Comparison between the richness obtained from the simulations and the analytical model for different values of the mutation rate $\mu$. The first row (a) and (b) corresponds to the infinite site model with no recombination and the second row (c) and (d) to the upper bound for two independent loci from \eq \ref{eq:phase3}. The left column corresponds to the total number of haplotypes and the second to the marginal distributions for the A and B alleles.}
\label{fig:phase1and3}
\end{center}
\end{figure}

Consider now a pair of loci A and B, and alleles in each locus. We assume that the mutation rates for each gene are low enough so that repeated mutations are rare (i.e. the infinite site model). When combining the two loci (no recombination occurs for now), they would simply behave like one long locus with mutation rate $\mu = \mu_{A} + \mu_{B}$ and, therefore, the first moment is simply $m_{0}(0) = - N (\mu_{A} + \mu_{B}) \ln{(\mu_{A} + \mu_{B})}$. We observe on \fig \ref{fig:phase1and3} (a) and (b) that simulations fit these results for the number of haplotypes and the marginal distributions for the A and B alleles. 

Let us now introduce a per generation recombination rate $r$. For a large enough population, the infinite site assumption holds, and one can consider recombination as another type of mutation. This yields an expected number of haplotypes with a maximum value of:

\begin{equation}
m_{0}(r)_{\infty} =- N (\mu_{A} + \mu_{B} + r) \ln{(\mu_{A} + \mu_{B} + r)} .
\label{eq:phase1}
\end{equation}

On the opposite extreme case of independent loci, the number of haplotypes in equilibrium can be computed too. It is essential to note that, if we were to have infinite mutations, the total number of alleles would be equal to the population since each allele would be distinct. In practice, such mutation rates do not happen, since an error threshold would occur at a finite mutation rate \cite{summers2006examining}. On the other hand, very high recombination rates can happen (up to 0.5 in the extreme case where one chooses randomly between two loci on different chromosomes). However, since recombination occurs between already existing haplotypes (unlike mutations that create new alleles), new combinations and, therefore, the number of haplotypes would be limited and dependent on the number of alleles and the mutation rates $\mu_{A}$ and $\mu_{B}$. The allele equilibrium distribution in both loci is not affected by recombination. To compute this upper bound, we now recombine the entire population without mutation as if there were infinite recombination. In practice, this corresponds to a Wright-Fisher process: we randomly choose two individuals and create their offspring with the allele A from the first parent and the allele B from the second parent. The two parents may have the same alleles (see the Appendix for derivations). We obtain:

\begin{equation}
m_{0 bound} = N^{2} \mu_{A}\mu_{B} \left[ \sum_{k = 1}^{N} \frac{e^{-\mu_{A} k}}{k} \ln{\left(1 + \frac{k}{\mu_{B}N}\right)} \right] .
\label{eq:phase3}
\end{equation}

Again, simulations in \fig \ref{fig:phase1and3} (c) and (d) confirm our results for the expected number of haplotypes and the marginal distributions for the A and B alleles. 

\begin{figure}[ht]
\begin{center}
\includegraphics[width=0.47\textwidth]{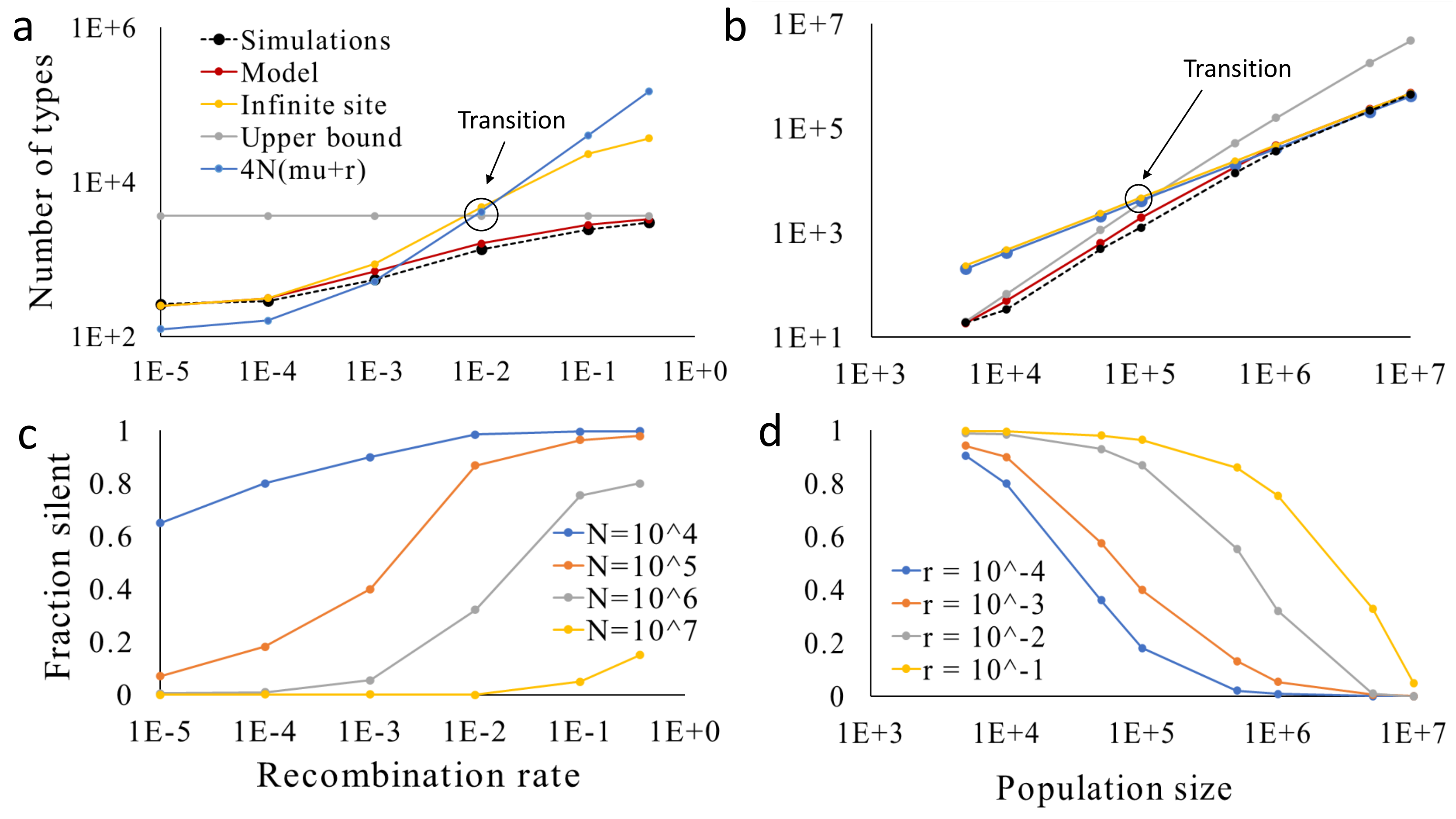}
\caption{Plots of the number of haplotypes as a function of the population size (a) and the recombination rate (b) and plots of the ratio of silent recombination as a function of the population size (c) and the recombination rate (d). The mutation rate is $10^{-4}$ for allele A and $2 \times 10^{-4}$ for allele B.}
\label{fig:phase2}
\end{center}
\end{figure}

The number of haplotypes will always be bounded by the lower of the two extreme cases. We denote the first regime as the ``infinite site" regime and the second one as the ``bounded" regime. For $N \gg 1/\mu$ (or tends to $\infty$) and $r$ small enough, the number of haplotypes from \eq \ref{eq:phase3} is higher than the one given by \eq \ref{eq:phase1} and, therefore, the maximum value for the number of haplotypes is given by the infinite site assumption with \eq \ref{eq:phase1}. For  $N \leq 1/ \mu$ and a high $r$, the opposite occurs and the maximum value is given by \eq \ref{eq:phase3}. In the infinite site regime, there are almost no silent recombination whereas, in the bounded regime, the ratio of silent recombination tends to 1 (depending on $r$) although never reaching 1 as observed  in \fig \ref{fig:phase2} (c) and (d).

For a given recombination rate, as $N$ increases, the fraction of silent recombination decreases from almost 1 to almost 0, and the number of haplotype shifts from the bounded regime to the infinite site regime. The higher the recombination rate, the larger $N$ needs to be for this transition to occur (\fig \ref{fig:phase2} (d)). For a given population size $N$, as $r$ increases, most haplotypes are created by recombination. As such, the fraction of silent recombination increases, and the number of haplotypes goes from the infinite site regime to the bounded regime (\fig \ref{fig:phase2} (c)).

Therefore, for intermediary values of $N$ (between $1/ \mu$ and $1000 / \mu$) and intermediary values of $r$ (between $\mu$ and 0.5), an intermediary regime emerges, and one can expect a mix between the two extreme regimes. We thus performed an interpolation where we compute a log regression with respect to $r$ between the value of the first moment for $r = \mu$ in \eq \ref{eq:phase1} and the value of the first moment at the upper bound from \eq \ref{eq:phase3}. 

\begin{equation}
m_{0}(r)_{interp} = 
\begin{cases}
- N(\mu_{A} + \mu_{B} + r) \ln{(\mu_{A} + \mu_{B} + r)} & r \leq \mu \\
\frac{\ln{r}}{\ln{\mu}} m_{0}(\mu) + \left( 1 - \frac{\ln{r}}{\ln{\mu}} \right) m_{0 bound} & r > \mu
\end{cases}
\label{eq:interp}
\end{equation}

The interpolation in \eq \ref{eq:interp} slightly overestimates $m_{0}$ in the bounded regime but it is much tighter to the simulations than the infinite site estimate as seen in \fig \ref{fig:phase2} (a) and (b) (see the Appendix and \cite{melka2020invasion} for a description of the simulations). 

Finally, since the upper bound from \eq \ref{eq:phase3} can be higher than the infinite site model for very large values of $N$ (after the transition), we need to take the minimum of this interpolation and the infinite site model in \eq \ref{eq:phase1}:

\begin{equation}
m_{0}(r) = \min{\left( m_{0}(r)_{interp} , m_{0}(r)_{\infty}\right)} .
\label{eq:final}
\end{equation}

In conclusion, given the first moment of the alleles (obtained from the marginal frequencies) and haplotypes frequencies, one can estimate $r$, as shall be further discussed. It already emerges that, for a given value of $m_{0}$ and an intermediary value of the population size, the recombination rate obtained from the infinite site model or Watterson's estimator is smaller than the actual one since it does not take into account the silent recombination.

\section{Recombination in the HLA Complex} 

To test the applicability of the boundary above to a real-life system, we analyzed the most polymorphic genes in the human genomes: the major histocompatibility complex (MHC). This locus is denoted as HLA (Human Leukocyte Antigen) in humans, on chromosome 6. This region is of interest since an HLA allele match between donors and recipients is crucial for recipient survival following solid organ or bone marrow transplants \cite{bradley1991role}. Given its importance, large-scale HLA typing of donors is performed by registries \cite{beatty1995impact, spellman2008advances, gragert2013six}. The fraction of patients in a population that can find appropriate donors depends on the frequency of their haplotypes \cite{slater2015power}. However, if the recombination rate is high, new haplotypes may be created too fast to allow full coverage.

The HLA gene complex contains the A, C, B, DR, and DQ genes, which together account for over 15,000 distinct alleles, and over a million haplotypes. Recent results suggest that a high haplotype creation rate could explain the observed haplotype polymorphism \cite{alter2017hla, lobkovsky2019multiplicative, simanovsky2019single}, in contrast with the low current recombination rates estimates \cite{blancher2012use}. However, current estimates do not incorporate silent recombination and, as such, may be underestimates. 

\begin{figure}[ht]
\begin{center}
\includegraphics[width=0.47\textwidth]{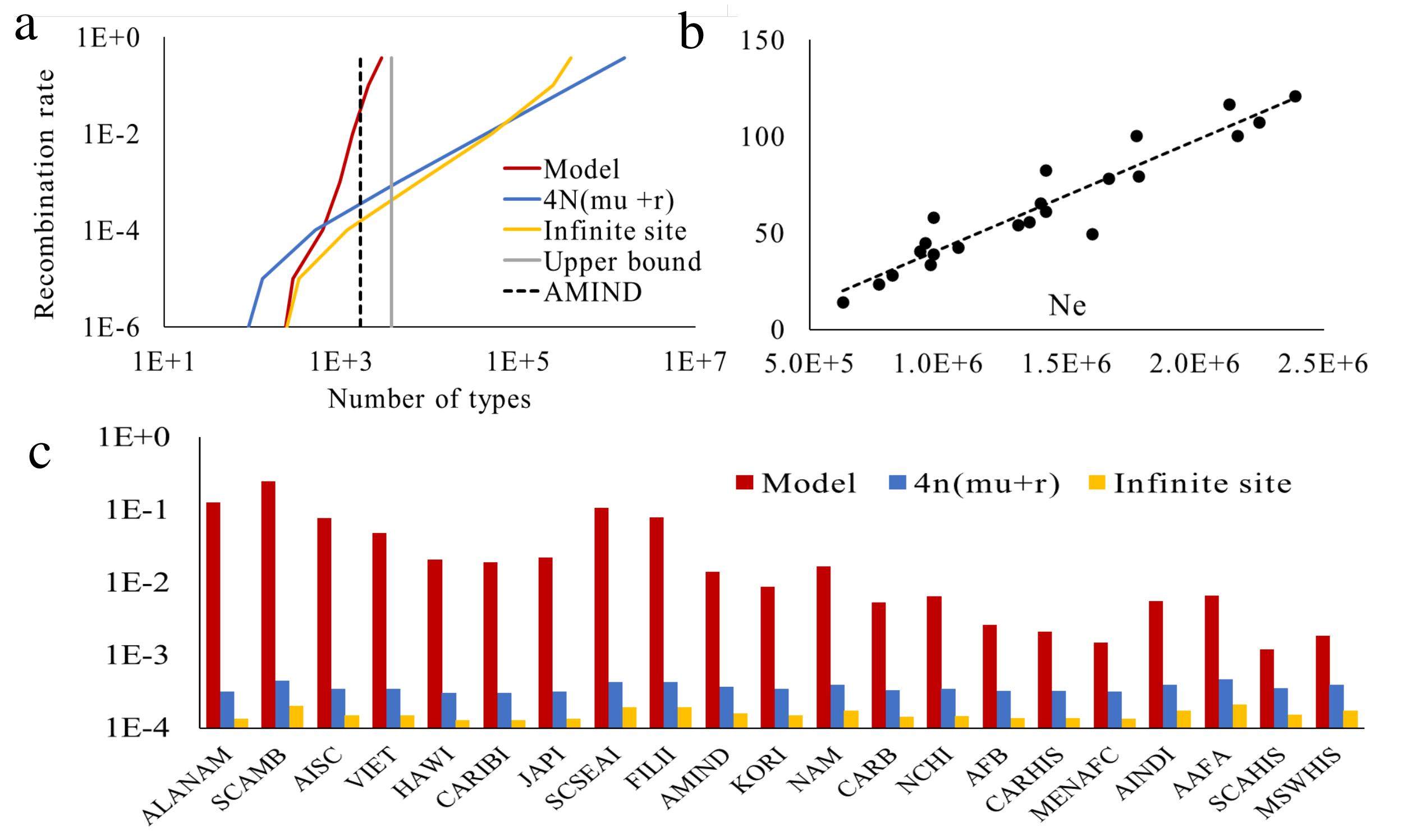}
\caption{Plot (a) - Recombination rate as a function of the number of haplotypes with different estimators. The red line corresponds to the number of haplotypes for AMIND (Amerindian population). We observe that our estimator yields a higher rate of recombination than the infinite site model or Watterson's estimator. Plot (b) - Ratio of the number of haplotypes for the A-C pair and the number of C alleles with respect to the effective population. We observe a linear relation. Plot (c) - Recombination rates across populations in the HLA region for the A-C pairs computed with our estimate, the infinite site model, and Watterson's estimator. Our estimator is from 10 to 1000 times higher.}
\label{fig:rates}
\end{center}
\end{figure}

To determine if one can expect a large number of silent recombination in the HLA locus, we analyzed the haplotype frequencies of 6.59 million donors from the National Marrow Donor Program registry, divided into 21 sub-populations \cite{israeli2021hla}. One can estimate $N_{e}$, using the marginal distributions of alleles, assuming an infinite site model for the number of alleles in each population. With a mutation rate of $1.45 \times 10^{-8}$ per base pair \cite{narasimhan2017estimating}, yielding an overall mutation rate of $\approx 8 \times 10^{-6}$ per gene (for a length of $\approx 550$ nucleotides per gene), we compute the effective population by inverting the richness formula from the Fisher log-series. Then, assuming an equal effective population for recombination and mutations, the recombination rate is computed by inverting \eq \ref{eq:final} as described in plot (a) in \fig \ref{fig:rates}. Plot (c) shows the recombination rates for the pair of genes A-C across populations (recombination rates for other pairs of genes may be found in the Appendix). Our estimate is 10 to 1000 times higher than the ones obtained from the infinite site model or Watterson's estimator, which is consistent with the presence of a large fraction of silent recombination.

Another relevant feature from our model is that the recombination rate is not linear with respect to the effective population. Instead, it has a square term as demonstrated in \eq \ref{eq:phase3} (obviously up to some bound). As the population grows, the total number of alleles and haplotypes grows. However, in addition to that, the fraction of non-silent recombination also increases with the population size inducing a second order correction term. This is in contrast with the number of mutations, which is linear with respect to the effective population (per definition in the current analysis). One can thus expect that, while the ratio between allele frequencies in different loci should be fixed among populations, the ratio between haplotypes and alleles should be linear in the effective population size. 

To validate this claim, we computed the ratio between the number of alleles (for instance, the ratio between the number of alleles in A over the number of alleles in C) and we computed the ratio between the number of haplotypes pairs and the number of alleles (for instance, the number of A-C pairs over the number of C alleles) as a function of the effective population size (\fig \ref{fig:rates} (c)). The latter ratio varies linearly with respect to the effective population, confirming that recombination does not behave like mutations (see the Appendix for a table with all regression coefficients between pairs).

\section{Conclusion}

The amount of genetic data and detailed haplotype samples have rapidly grown over the last few years. Nevertheless, precise methods to use such samples in order to estimate the recombination rate within haplotypes are still lacking.

We have here proposed a new estimate of the number of haplotypes that incorporates the difference between recombination and mutations. Recombination draws from a pool of existing alleles, some very frequent, and, as such, quite often, reproduces existing haplotypes. The resulting number of haplotypes is bounded at a level much lower than the total population even for a very high recombination rate. To the best of our knowledge, silent recombination was seldom considered when estimating the recombination rate. 

Nevertheless, our estimator, although more accurate than existing models, suffers a few caveats. For the computation of the mutation rate for each allele, we use a neutral infinite site model as is most standard in genetic research, but this may not be the case in all loci \cite{alter2017hla}. Such a deviation could be due, as is classically argued, to selection \cite{hedrick1983evidence}, or might be the results of other mechanisms, such as catastrophes \cite{melka2020invasion}. For example, in the HLA locus studied here, there is a disparity between the computed and actual distribution (see the Appendix). Besides, we assumed that allele distribution was at equilibrium in order to compute our upper bound on the number of haplotypes, but a very long time is required to reach such an equilibrium, especially for low mutation rates. Finally, the flattening slope of our estimate close to the upper bound might result in large differences in the recombination rate even for small differences in the number of haplotypes. This problem is aggravated if the allele equilibrium is not achieved.

We have here studied a purely neutral model with the type size distribution of a Moran model. Obviously, in the presence of selection, the allele size distribution would differ, and, accordingly, the recombination rate estimate. Such effects will be even more important with epistatic selection, where recombination and selection interfere \cite{neher2009competition, weissman2012limits, held2019survival}. Still, in the presence of selective sweeps, the fraction of individuals in very large families will actually increase, and, with it, the fraction of silent recombination.

Recombination rates estimates are of interest in the HLA locus, where, although haplotype frequencies are estimated over very large populations, the within haplotype recombination rate is still debated. Most current recombination rate estimates use coalescent models and Watterson's estimator. We analyzed data from the HLA locus and obtained a ten to a thousand-time higher recombination rate than currently estimated. This difference and the non-linearity of the number of haplotypes with respect to the effective population size are evidence that recombination cannot be treated as another type of mutation due to the presence of silent recombination. This high haplotype creation rate is in agreement with recent results \cite{lobkovsky2019multiplicative}, and it implies that, unless huge surveys are conducted, genome registries will seldom approach an exhaustive list of existing haplotypes.

\appendix

\section{Results derivations}

\subsection{Mutation model}

In a population with different alleles, we define as $P_{k}$, the probability for an allele to be of size $k$. We define $\alpha$ as the birth and death rate (which we assume equal so that the population stays constant) and can be set equal to 1 up to a time scaling. $\mu$ is the per generation mutation rate. A type of size $k$ can endure a death at rate $\alpha$ or a birth at rate $\alpha(1 - \mu)$ and not be of size $k$ anymore. A type of size $k-1$ can have a birth at rate $\alpha(1 - \mu)$ and become a type of size $k$. Finally, a type of size $k + 1$ can die at rate $\alpha$ and become a type of size $k$. The dynamics of $P_{k}$ is:

\begin{equation}
\frac{dP_{k}}{dt} = \alpha \left[ -(2 - \mu) kP_{k} + (1-\mu)(k-1)P_{k-1} + (k+1)P_{k+1} \right]
\end{equation}

In steady state, $\frac{dP_{k}}{dt} = 0$. Also, we assume $P_{k}$ to be a smooth enough function (in the infinite site model), so, if we denote $P_{k} = P$, we get:

\begin{equation}
\begin{aligned}
P_{k + 1} &= P + P' + 1/2P" \\
P_{k - 1} &= P - P' + 1/2P"
\end{aligned}
\end{equation}

\begin{equation}
\begin{aligned}
0 & = -2kP  + k\mu P + (k-1)(1 - \mu)(P - P' + 1/2 P'') \\
	&+ (k+1)(P + P' + 1/2 P'') \\
0 & = \underbrace{kP" + 2P'}_{(kP)"} + \underbrace{\mu kP' + \mu P}_{\mu(kP)'} \underbrace{ - \mu P' - 1/2\mu (k-1)P"}_{\text{negligible}} \\
0 & = (kP)" + \mu(kP)'
\end{aligned}
\end{equation}

We define $Q = kP$ and get:

\begin{equation}
\begin{aligned}
Q'' + \mu Q' &= 0 \\
Q' + \mu Q & = B \\
(Qe^{\mu k})' &= Be^{\mu k} \\
Qe^{\mu k} &= \frac{B}{\mu}e^{\mu k} + A \\
Q &= \frac{B}{\mu} + Ae^{-\mu k} \\
Q &= Ae^{-\mu k}
\end{aligned}
\end{equation}

$B = 0$ because of the limit condition $\lim_{k \rightarrow + \infty} P_{k} = 0$. So, we finally get:

\begin{equation}
P_{k} = \frac{A}{k} e^{-\mu k}
\label{eq:proba}
\end{equation}

This result is actually the Fisher log-series  obtained in a simplified way. We define as $\bar{n}$ the average size of a family, $N$ the total population and therefore the average number of types or first moment $m_{0} = N / \bar{n}$. $\mu$ is considered small in order to make approximations.

\begin{equation}
\bar{n} = \sum \limits_{k=1}^{\infty} k P_{k} = A  \sum \limits_{k=1}^{\infty} e^{-\mu k} = \frac{A e^{-\mu}}{1 - e^{-\mu}} \approx \frac{A(1 - \mu)}{\mu} \approx \frac{A}{\mu}
\end{equation}

\begin{equation}
\begin{aligned}
1 &= \sum \limits_{k=1}^{\infty} P_{k} = A \sum \limits_{k=1}^{\infty} \frac{ e^{-\mu k}}{k} \\
1 &= A \sum \limits_{k=1}^{\infty} \left[ \int_{\mu}^{\infty}  e^{-\theta k} d\theta \right] = A \int_{\mu}^{\infty} \left[ \sum \limits_{k=1}^{\infty}  e^{-\theta k} \right] d\theta \\
1 &= A \int_{\mu}^{\infty} \frac {e^{-\theta}}{1 -  e^{-\theta}} d\theta = A \int_{e^{-\mu}}^{0} \frac{-x}{1-x} \frac{dx}{x} = A \int^{e^{-\mu}}_{0} \frac{1}{1-x} dx \\
1 &= - A \left[ \ln (1-x) \right]_{0}^{e^{-\mu}} = - A \ln(1 - e^{-\mu}) \approx -A \ln{\mu}
\end{aligned}
\label{eq:sum}
\end{equation}

\begin{equation}
\implies m_{0} = \frac{N}{\bar{n}} \approx - N\mu \ln{\mu}
\label{eq:mu}
\end{equation}

\begin{equation}
P_{k} = - \frac{e^{-\mu k}}{k \ln{\mu}}, \>\>  N_{k} = m_{0} P_{k} = \frac{N \mu e^{-\mu k}}{k}
\label{eq:proba2}
\end{equation}

\subsection{Two-locus model with no recombination}

In this regime, we assume asexual reproduction and an ``infinite site" model (i.e. each mutation leads to the creation of a new type). The mutation rate in gene A is $\mu_{A}$ and in gene B is $\mu_{B}$. A birth event leads to 4 possible outcomes:

\begin{enumerate}
\item{} no mutation with probability $(1 - \mu_{A})(1 - \mu_{B})$,
\item{} A mutation on gene A with probability $\mu_{A}(1 - \mu_{B})$,
\item{} A mutation on gene B with probability $\mu_{B}(1 - \mu_{A})$,
\item{} A mutation on both genes with probability $\mu_{A}\mu_{B}$.
\end{enumerate}

One individual is randomly selected. In case of no mutation, the corresponding type size is merely increased by 1. In the case of mutation on gene A, a new type is created with size 1. Its A allele is new, and its B allele is the same as the original individual, and vice versa for mutations in B. In the case of a double mutation, both alleles are new. Since genes A and B are independent, we can treat their combination as a single gene, and the mutation rate is, therefore, $\mu = \mu_{A} + \mu_{B}$. We already solved this model and obtained the probability $P_{k}$ for a type to have a size $k$ is given by:

\begin{equation}
P_{k} = - \frac{e^{-(\mu_{A} + \mu_{B}) k}}{k \ln{(\mu_{A} + \mu_{B})}},
\end{equation}

leading to a total number of types

\begin{equation}
m_{0} \approx - N(\mu_{A} + \mu_{B}) \ln({\mu_{A} + \mu_{B}})
\end{equation}

\subsection{Infinite recombination model}

\begin{figure}[h]
\begin{center}
\includegraphics[width=0.45\textwidth]{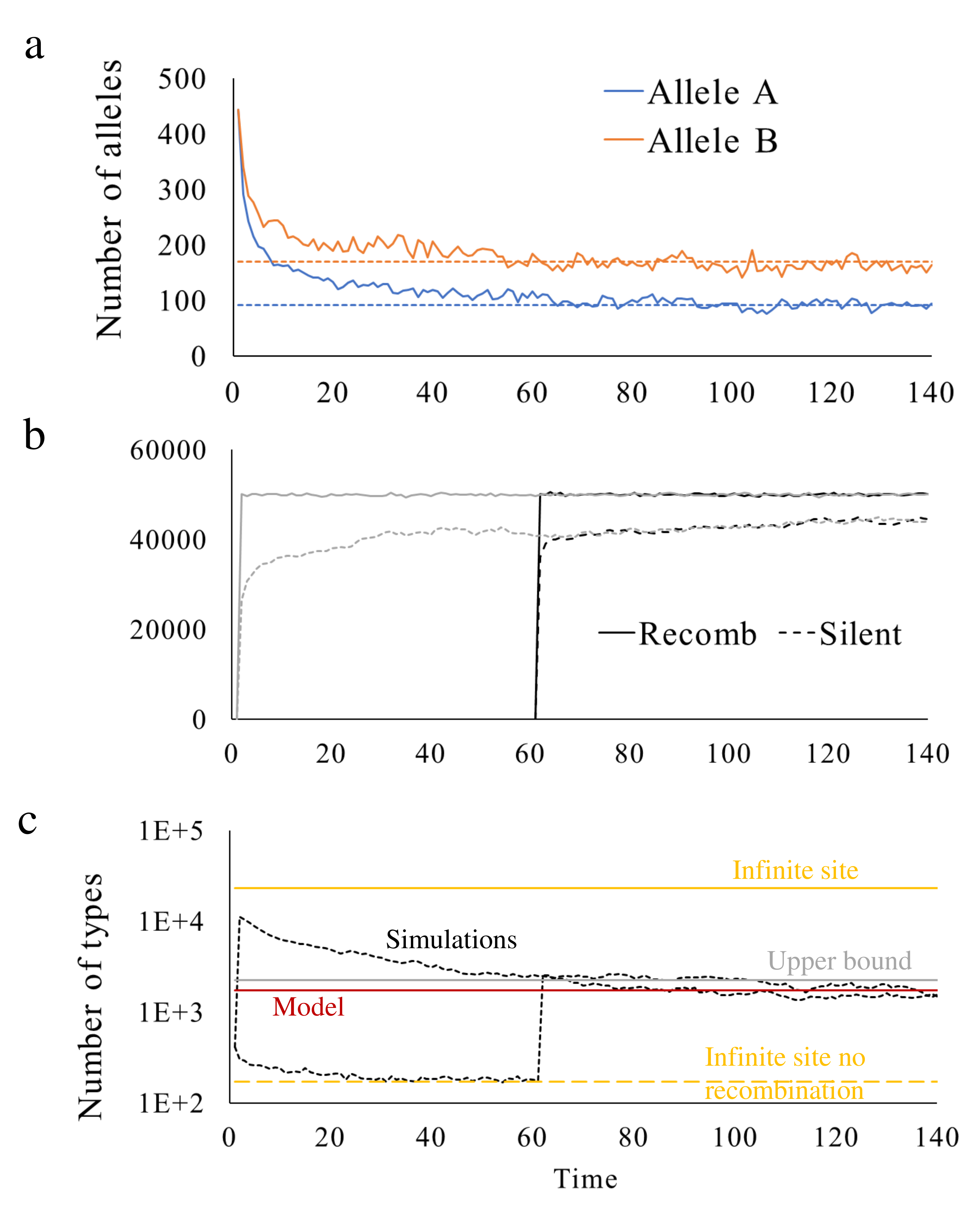}
\caption{Time series of the simulations. The final equilibrium is the same whether the recombination process started from time 0 or after a lag. The mutation rate is $10^{-4}$ for The A alleles and $2 \times 10^{-4}$ for the B alleles. The number of steps is $1.4 \times 10^{9}$. The recombination rate is $0.1$. Plot (a) represents the time series for the the number of alleles for genes A and B. Plot (b) represents the time series of the number of recombination and silent recombination. Plot (c) represents time series of the simulations of the number of types.}
\label{fig:convergence}
\end{center}
\end{figure}

In this regime, we assume sexual reproduction and the allele distribution for the two genes are at equilibrium according to the previous model. We randomly select two existing individuals (they can belong to the same type). the newly created individual has the same A allele as the first selected individual and the same B allele as the second. The probability of choosing a type $(A_{i}, B_{j})$ with size $N_{i,j}$ is $p_{i,j} = \frac{N_{i,j}}{N}$. The probability of choosing an individual with a given allele $A_{i}$ is $p_{i} = \sum_{j}\frac{N_{i,j}}{N}$. Similarly, the probability of choosing an individual with allele $B_{j}$ is $q_{j} = \sum_{i}\frac{N_{i,j}}{N}$. There are therefore $Np_{i}$ individuals with allele $A_{i}$ and out of those, $Np_{i}q_{j}$ individuals of type $(A_{i}, B_{j})$. This implies that $N_{i,j} = Np_{i}q_{j}$ and $p_{i,j} = p_{i}q_{j}$. We assume that steady state is achieved from the first regime, and A and B alleles  are distributed accordingly. This assumption is consistent as seen in \fig \ref{fig:convergence}. Indeed, simulations where we wait for the alleles to achieve equilibrium or start the recombination process from the beginning yield the same number of haplotypes. We then determine, for each individual, if their A allele is $i$ and if their B allele is $j$. $\mathbb{E}[N(k)] = \sum_{i,j} \mathbb{P}(N_{i,j} = k)$ is the expected number of types with size $k$ with $\mathbb{P}(N_{i,j} = k) = {N\choose k} p_{i,j}^{k} (1 - p_{i,j})^{N - k}$,

\begin{widetext}
\begin{equation}
\begin{aligned}
\mathbb{E}[N(k)]		&= {N\choose k} \sum_{i,j} (p_{i} q_{j})^{k} (1 - p_{i} q_{j})^{N - k} \\
				&\approx {N\choose k} \sum_{k_{A},k_{B}} N(k_{A}) N(k_{B}) \left(\frac{k_{A}}{N} \frac{k_{B}}{N} \right)^{k} \left(1 - \frac{k_{A}}{N} \frac{k_{B}}{N} \right)^{N - k} \\
				&= {N\choose k} m_{0_{A}} m_{0_{B}} \sum_{k_{A},k_{B}} P_{k_{A}} P_{k_{B}} \left(\frac{k_{A}}{N} \frac{k_{B}}{N} \right)^{k} \left(1 - \frac{k_{A}}{N} \frac{k_{B}}{N} \right)^{N - k} \\
				&= {N\choose k} (N \mu_{A} \ln (\mu_{A})) (N \mu_{B} \ln (\mu_{B})) \sum_{k_{A},k_{B}} \frac{e^{-\mu_{A} k_{A}}}{k_{A} \ln{(\mu_{A})}} \frac{e^{-\mu_{B} k_{B}}}{k_{B} \ln{(\mu_{B})}} \left(\frac{k_{A}}{N} \frac{k_{B}}{N} \right)^{k} \left(1 - \frac{k_{A}}{N} \frac{k_{B}}{N} \right)^{N - k} \\
				&= {N\choose k} N^{2} \mu_{A}\mu_{B} \sum_{k_{A},k_{B}} \frac{e^{-\mu_{A} k_{A}}}{k_{A}} \frac{e^{-\mu_{B} k_{B}}}{k_{B}} \left(\frac{k_{A}}{N} \frac{k_{B}}{N} \right)^{k} \left(1 - \frac{k_{A}}{N} \frac{k_{B}}{N} \right)^{N - k} \\
\end{aligned} 
\end{equation}
\end{widetext}

\begin{widetext}
\begin{equation}
\begin{aligned}
\mathbb{E}[N(1)]		&= N N^{2} \mu_{A}\mu_{B} \sum_{k_{A},k_{B}} \frac{e^{-\mu_{A} k_{A}}}{k_{A}} \frac{e^{-\mu_{B} k_{B}}}{k_{B}} \left(\frac{k_{A}}{N} \frac{k_{B}}{N} \right)^{1} \left(1 - \frac{k_{A}}{N} \frac{k_{B}}{N} \right)^{N - 1} \\
				&= N \mu_{A}\mu_{B} \sum_{k_{A},k_{B}} e^{-\mu_{A} k_{A}} e^{-\mu_{B} k_{B}} \underbrace{\left(1 - \frac{k_{A}}{N} \frac{k_{B}}{N} \right)^{N - 1}}_{\approx e^{-\frac{k_{A}k_{B}}{N}}} \\
				&=  N \mu_{A}\mu_{B} \sum_{k_{A}} e^{-\mu_{A} k_{A}}  \sum_{k_{B}} e^{- k_{B} \left( \mu_{B} + \frac{k_{A}}{N} \right)} =  N \mu_{A}\mu_{B} \sum_{k} \frac{e^{- \mu_{B} - k \left( \mu_{A} + \frac{1}{N} \right)}}{1 - e^{- \left( \mu_{B} + \frac{k}{N} \right)}} \\
\end{aligned} 
\end{equation}
\end{widetext}

\begin{widetext}
\begin{equation}
\begin{aligned}
m_{0}		&= \sum_{k} \mathbb{E}[N(k)] \\
		&= \sum_{k} {N\choose k} N^{2} \mu_{A}\mu_{B} \sum_{k_{A},k_{B}} \frac{e^{-\mu_{A} k_{A}}}{k_{A}} \frac{e^{-\mu_{B} k_{B}}}{k_{B}} \left(\frac{k_{A}}{N} \frac{k_{B}}{N} \right)^{k} \left(1 - \frac{k_{A}}{N} \frac{k_{B}}{N} \right)^{N - k} \\
		&= N^{2} \mu_{A}\mu_{B} \sum_{k_{A},k_{B}} \frac{e^{-\mu_{A} k_{A}}}{k_{A}} \frac{e^{-\mu_{B} k_{B}}}{k_{B}} \underbrace{\left[ \sum_{k} {N\choose k} \left(\frac{k_{A}}{N} \frac{k_{B}}{N} \right)^{k} \left(1 - \frac{k_{A}}{N} \frac{k_{B}}{N} \right)^{N - k} \right]}_{\underbrace{1^{N} - \left(1 - \frac{k_{A}}{N} \frac{k_{B}}{N} \right)^{N}}_{\approx 1 - e^{-\frac{k_{A}k_{B}}{N}}}}\\
		&= N^{2} \mu_{A}\mu_{B} \left[ \sum_{k_{A}} \sum_{k_{B}} \left[ \frac{e^{-\mu_{A} k_{A}}}{k_{A}} \frac{e^{-\mu_{B} k_{B}}}{k_{B}} - \frac{e^{-\mu_{A} k_{A}}}{k_{A}} \frac{e^{-\mu_{B} k_{B}}}{k_{B}} e^{-\frac{k_{A}k_{B}}{N}} \right] \right] \\
		&=  N^{2} \mu_{A}\mu_{B} \left[ \sum_{k_{A}} \sum_{k_{B}} \frac{e^{-\mu_{A} k_{A}}}{k_{A}} \frac{e^{-\mu_{B} k_{B}}}{k_{B}} - \sum_{k_{A}} \sum_{k_{B}} \frac{e^{-\mu_{A} k_{A}}}{k_{A}} \frac{e^{- k_{B} \left( \mu_{B} + \frac{k_{A}}{N} \right) }}{k_{B}} \right] \\
		&=  N^{2} \mu_{A}\mu_{B} \left[ \underbrace{ \left( \sum_{k_{A}} \frac{e^{-\mu_{A} k_{A}}}{k_{A}} \right)}_{\approx -\ln{\mu_{A}}} \underbrace{ \left( \sum_{k_{B}} \frac{e^{-\mu_{B} k_{B}}}{k_{B}} \right)}_{\approx -\ln{\mu_{B}}} - \sum_{k_{A}} \frac{e^{-\mu_{A} k_{A}}}{k_{A}} \underbrace{\left( \sum_{k_{B}} \frac{e^{- k_{B} \left( \mu_{B} + \frac{k_{A}}{N} \right) }}{k_{B}} \right) }_{\approx - \ln{\left( \mu_{B} + \frac{k_{A}}{N}\right)}} \right] \\
		&= N^{2} \mu_{A}\mu_{B} \left[ \ln{\mu_{A}} \ln{\mu_{B}} + \sum_{k_{A}} \frac{e^{-\mu_{A} k_{A}}}{k_{A}} \ln{\left( \mu_{B} + \frac{k_{A}}{N}\right)} \right] \\
		&= N^{2} \mu_{A}\mu_{B} \left[ \ln{\mu_{A}} \ln{\mu_{B}} + \sum_{k_{A}} \frac{e^{-\mu_{A} k_{A}}}{k_{A}} \left[ \ln{\mu_{B}} + \ln{\left(1 + \frac{k_{A}}{\mu_{B}N}\right)} \right]  \right] \\
		&=  N^{2} \mu_{A}\mu_{B} \left[ \ln{\mu_{A}} \ln{\mu_{B}} + \ln{\mu_{B}} \underbrace{ \left( \sum_{k_{A}} \frac{e^{-\mu_{A} k_{A}}}{k_{A}} \right)}_{\approx -\ln{\mu_{A}}} + \sum_{k_{A}} \frac{e^{-\mu_{A} k_{A}}}{k_{A}} \ln{\left(1 + \frac{k_{A}}{\mu_{B}N}\right)} \right] \\
		&= N^{2} \mu_{A}\mu_{B} \left[ \sum_{k} \frac{e^{-\mu_{A} k}}{k} \ln{\left(1 + \frac{k}{\mu_{B}N}\right)} \right]
\label{eq:m0(1)}
\end{aligned}
\end{equation}
\end{widetext}

In \eq \ref{eq:m0(1)}, the expected number of haplotypes is actually different from $N^{2} \mu_{A} \mu_{B} \ln{(\mu_{A})} \ln{(\mu_{B})}$, which represents the product of all combinations of A and B alleles, since, even in the steady state, not all possible pairs of alleles are present.

Just as in \eq \ref{eq:sum}, the finite sum can be approximated by the logarithm since $\mu \ll 1$ and $N$ is large so the remainder of the sum is negligible.

\begin{figure}[h]
\begin{center}
\includegraphics[width=0.45\textwidth]{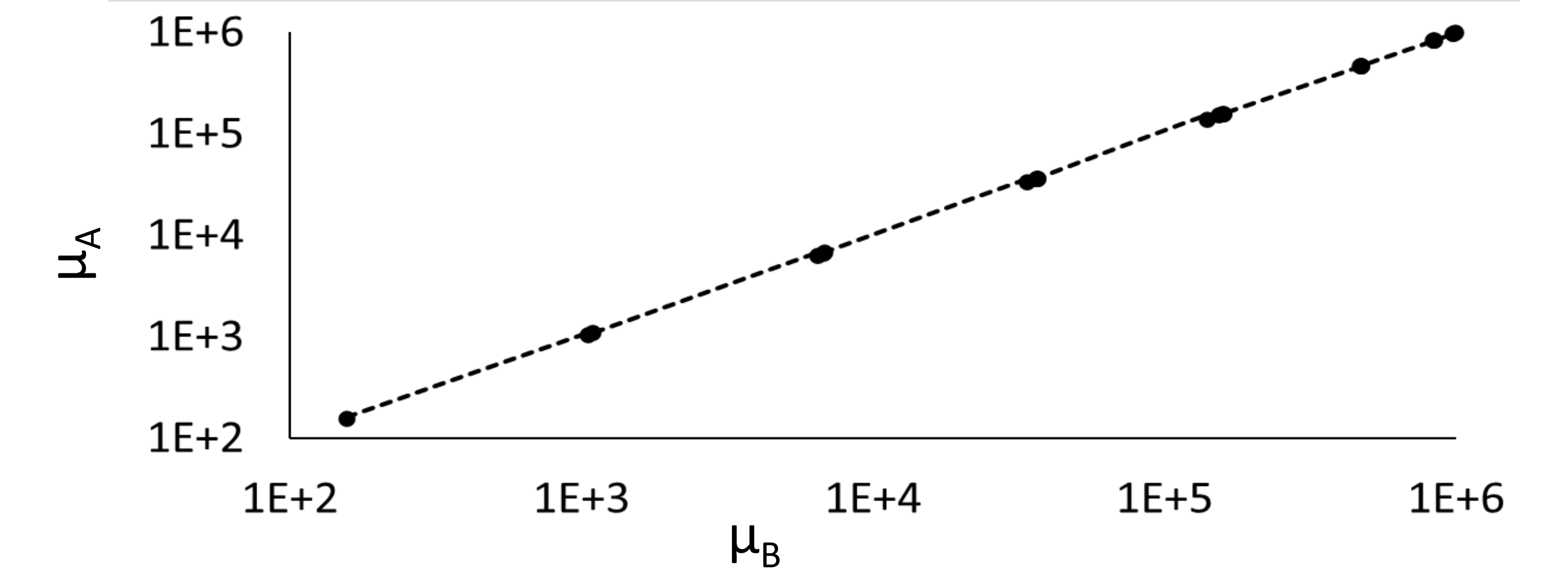}
\caption{Computation of the expected number of haplotypes in the infinite recombination model for different values of $\mu_{A}$ and $\mu_{B}$ (dots). The x-axis represents $m_{0}$ from \eq \ref{eq:m0(1)} and the y-axis is obtained when inverting $\mu_{A}$ and $\mu_{B}$. The dots are on the first diagonal confirming that \eq \ref{eq:m0(1)} is symmetric.}
\label{fig:symmetry}
\end{center}
\end{figure}

Although \eq \ref{eq:m0(1)} does not seem to be symmetric with respect to gene A and B due to the approximations we performed in the computation, inverting $\mu_{A}$ and $\mu_{B}$ would yield almost identical results as can be observed in \fig \ref{fig:symmetry}.

\subsection{Intermediary regime}

\subsubsection{Lower bound interpolation}

We tested whether a regular birth and death process with mutations and recombination could be modeled as a combination of the 2 regimes described above and if the ratio between those 2 regimes is linked to the recombination rate $r$. We approximate that the creation of new types (with a size of 1) comes from two sources: either the recombination regime or the no recombination regime. The extinction of types of size 1 comes from the mixed regime. At equilibrium, we assume creations and extinctions are equal. We denote by 0 the no recombination regime, by 1 the infinite recombination regime, and by $r$ the mixed regime. $P_{1}$ is the probability for a type of size 1, $m_{0}$ is the number of types, and $m_{1}$ is the total population. We obtain:

\begin{equation}
m_{0}(0) = - N(\mu_{A} + \mu_{B}) \ln({\mu_{A} + \mu_{B}})
\end{equation}

\begin{equation}
P_{1}(0) = - \frac{ e^{-(\mu_{A} + \mu_{B})}}{\ln{(\mu_{A} + \mu_{B})}}
\end{equation}

\begin{equation}
m_{0}(1) = N^{2} \mu_{A}\mu_{B} \left[ \sum_{k = 1}^{\infty} \frac{e^{-\mu_{A} k}}{k} \ln{\left(1 + \frac{k}{\mu_{B}N}\right)} \right]
\end{equation}

\begin{equation}
P_{1}(1) = \frac{\mathbb{E}[N(1)]}{m_{0}(1)}
\end{equation}

The number of creations of new types is given by:

\begin{equation}
r \frac{m_{0}(1) P_{1}(1)}{m_{1}} + (1 - r) \frac{m_{0}(0) P_{1}(0)}{m_{1}}
\end{equation}

The number of extinctions of types of size 1 is given by:

\begin{equation}
\frac{P_{1}(r) m_{0}(r)}{m_{1}}
\end{equation}

At equilibrium, creations and extinctions rates are equal. This yields:

\begin{equation}
m_{0}(r)_{int1} = \frac{r m_{0}(1) P_{1}(1) + (1 - r) m_{0}(0) P_{1}(0)}{P_{1}(r)}
\end{equation}

We also assume that $P_{1}(r)$ is a linear combination of $P_{1}(0)$ and $P_{1}(1)$: 

\begin{equation}
P_{1}(r) = (1 - r) P_{1}(0) + r P_{1}(1)
\end{equation}

\subsubsection{Upper bound interpolation}

We also developed another interpolation between the two extreme regimes of no recombination and infinite recombination. In this version, we simply compute a log regression between the value of $m_{0}$ for $r = \mu$ given by $m_{0}(\mu) = - 2N(\mu_{A} + \mu_{B}) \ln{(2(\mu_{A} + \mu_{B}))}$ and $m_{0}(1)$ from \eq \ref{eq:m0(1)}. This yields:

\begin{equation}
m_{0}(r)_{int2} =  
\begin{cases}
- N(\mu_{A} + \mu_{B} + r) \ln{(\mu_{A} + \mu_{B} + r)} & r \leq \mu \\
\frac{\ln{r}}{\ln{\mu}} m_{0}(\mu) + \left( 1 - \frac{\ln{r}}{\ln{\mu}} \right) m_{0}(1) &  r > \mu
\end{cases}
\end{equation}

\subsubsection{Mixed model}

The upper interpolation is quite a good fit to our simulation. Nevertheless, in some cases it slightly overestimates the actual number of types $m_{0}$. Since the first interpolation slightly underestimates $m_{0}$, we took the average of those two and achieve an even better fit from our simulations. This average, although heuristic, provides a more accurate estimate of the first moment since both aforementioned interpolations are actually close to each other compared to the upper bounds. An intermediate estimate is therefore, a natural choice and the average is the simplest one. We tested other ratios and yield similar results.

\begin{equation}
m_{0}(r) = \frac{m_{0}(r)_{int1} + m_{0}(r)_{int2}}{2}
\label{eq:int}
\end{equation}

\section{Simulations}

In the simulations, the population is composed of several haplotypes resulting from the combinations of the alleles from genes A and B. We assume birth and death rates to be equal so that the population is constant. For simplicity, at each step, a birth and a death event occur. Each birth results either in a regular birth where a haplotype simply increases its size by 1, or a mutation in A or B alleles or in both, hence creating a new type of size 1, or finally, recombination where the offspring gets its A allele  from one parent and its B allele  from the other. We compute the normalized probabilities for all those events and randomly choose which will occur. For the sake of efficiency, all the initial haplotypes (and thereafter all haplotypes) are plugged into a tree, in order to keep track of each haplotype size. Each leaf corresponds to a haplotype and the number associated with this entry is the haplotype size. The value of each internal node in the tree is the sum of its two sons, the tree root being the size of the total population.  We run those simulations for a number of steps sufficiently large so that we achieve a steady state.

\section{Effective population}

Knowing the actual number of haplotypes $m_{0}$ for a population, we can numerically invert \eq \ref{eq:int} and compute $r$. We first need to compute the effective population $N_{e}$ by inverting \eq \ref{eq:mu}. Therefore, we also need $\mu$ independently of $N_{e}$. One way to do it is to use the probabilities $P_{k}$ from \eq \ref{eq:proba2}.

We assume an infinite site model, $P_{k}$ is continuous. We design bins around $k$ with length $\Delta_{k}$. Therefore, the number of alleles $N_{k}$ with a size comprised in that bin is:

\begin{equation}
N_{k} = P_{k} \Delta_{k} m_{0}
\end{equation}

Taking the log on both sides and using \eq \ref{eq:proba}, we get:

\begin{equation}
- \ln{\left[\frac{k N_{k}}{\Delta_{k} m_{0}}\right]} = \mu k + \ln{\left[\left| \ln{(\mu)} \right| \right]}
\end{equation}

Since for each population, we have the frequencies of all the haplotypes, we determine the marginal distribution for a given gene and compute $N_{k}$ for each bin. We then perform a linear regression to get $\mu$.

\begin{figure}[h]
\begin{center}
\includegraphics[width=0.45\textwidth]{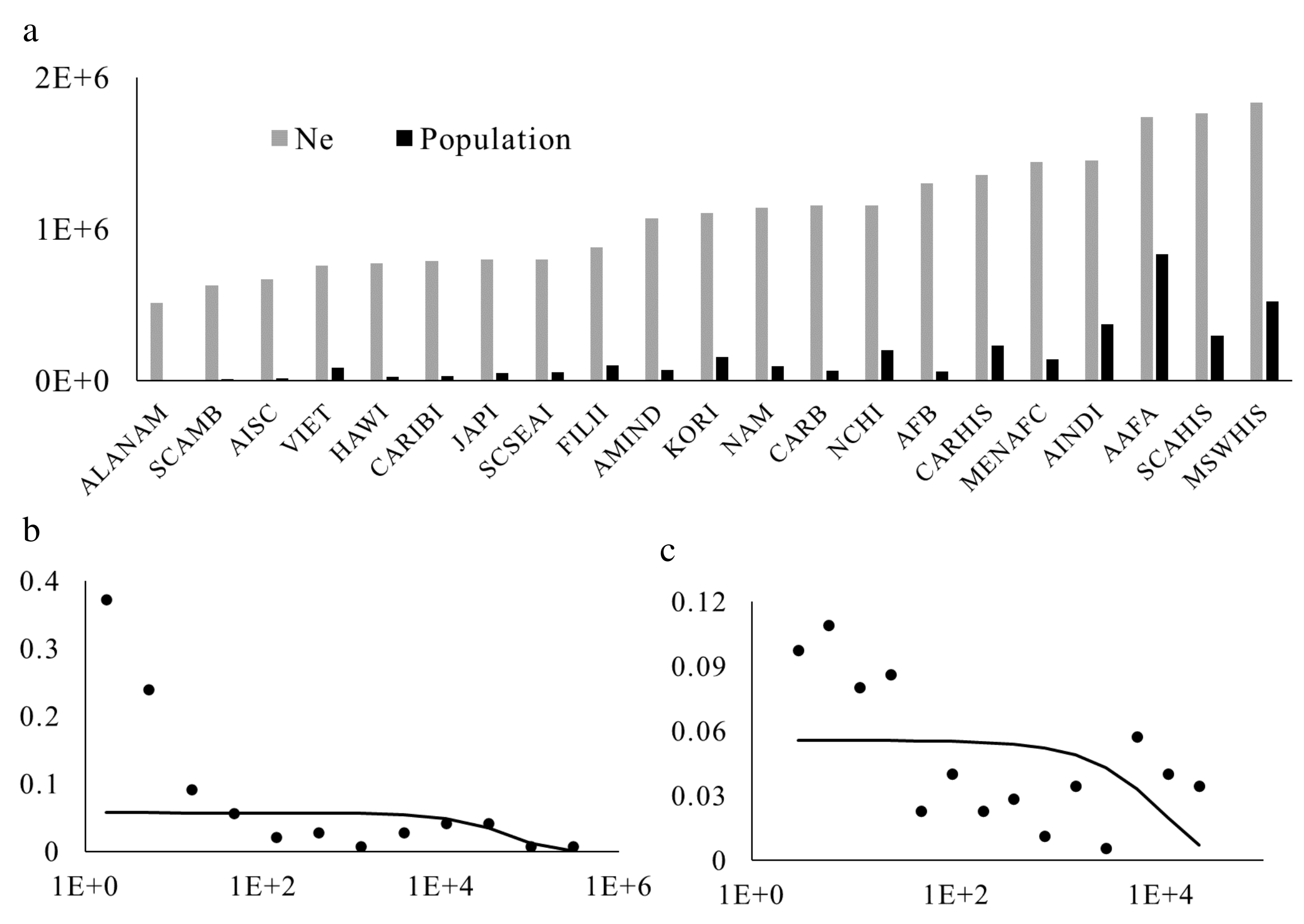}
\caption{Plot (a) represents the respective population and effective population. Distribution of the alleles (dots). The full line represents the theoretical distribution with $\mu = 8 \times 10^{-6}$. Plot (b) is for the gene C in the HIS population. Plot (c) is for the gene B in the KORI population.}
\label{fig:fattails}
\end{center}
\end{figure}

One problem we encountered is that the actual distributions have fat tails (power law distributions with an exponential cutoff) and, therefore, deviate from our estimate as in \fig \ref{fig:fattails} plots (b) and (c). Thus, we use the mutation rate from an external source computed using statistics on the genes. We get the effective populations in plot (a).

\section{HLA complex}

We demonstrated that recombination can be assumed to behave like mutations only for low rates and large populations. For mutations, the number of alleles is linear with respect to the effective population. If we assume that the effective population is the same for each gene, then the ratio between the number of alleles in two loci should be constant. On the other hand, according to our estimate, the number of haplotypes is not linear in $N_{e}$. Therefore, the ratio of the number of haplotypes and alleles varies among populations. \tab \ref{tab:reg} represents the slope coefficients for linear regression and their p-values. The slopes for the ratio of the number of alleles are close to 0 with high p-values, confirming that those ratios do not depend on $N_{e}$, whereas the slopes for the ratio of haplotypes over alleles are different from 0 with very small p-values. 

\begin{table}
\caption{Linear regression coefficients between gene pair and haplotype ratios.}
\begin{ruledtabular}
\begin{tabular}{lcc}
Ratio	 	&Slope 	&p-value	\\ 
\hline
$m_{0_{A}} / m_{0_{C}}$ 		& 0.2286 	& 0.1236 		\\
$m_{0_{C}} / m_{0_{B}}$  		& -0.3717	& 7.1953E-2 		\\
$m_{0_{B}} / m_{0_{DR}}$ 		& -0.2187 	& 0.2255 		\\
$m_{0_{DR}} / m_{0_{DQ}}$ 		& 0.74615 	& 1.9447E-6 		\\
\hline
$m_{0_{A,C}} / m_{0_{C}}$  		& 0.4270 	& 2.8047E-2 		\\
$m_{0_{C,B}} / m_{0_{B}}$  		& 0.5715 	& 1.1130E-3	 	\\
$m_{0_{B,DR}} / m_{0_{DR}}$  		& 0.6226 	& 1.2045E-5 		\\
$m_{0_{DR,DQ}} / m_{0_{DQ}}$  	& 0.9334 	& 4.3806E-14 	\\ 
\end{tabular}
\end{ruledtabular}
\label{tab:reg}
\end{table}

\begin{figure}[h]
\begin{center}
\includegraphics[width=0.45\textwidth]{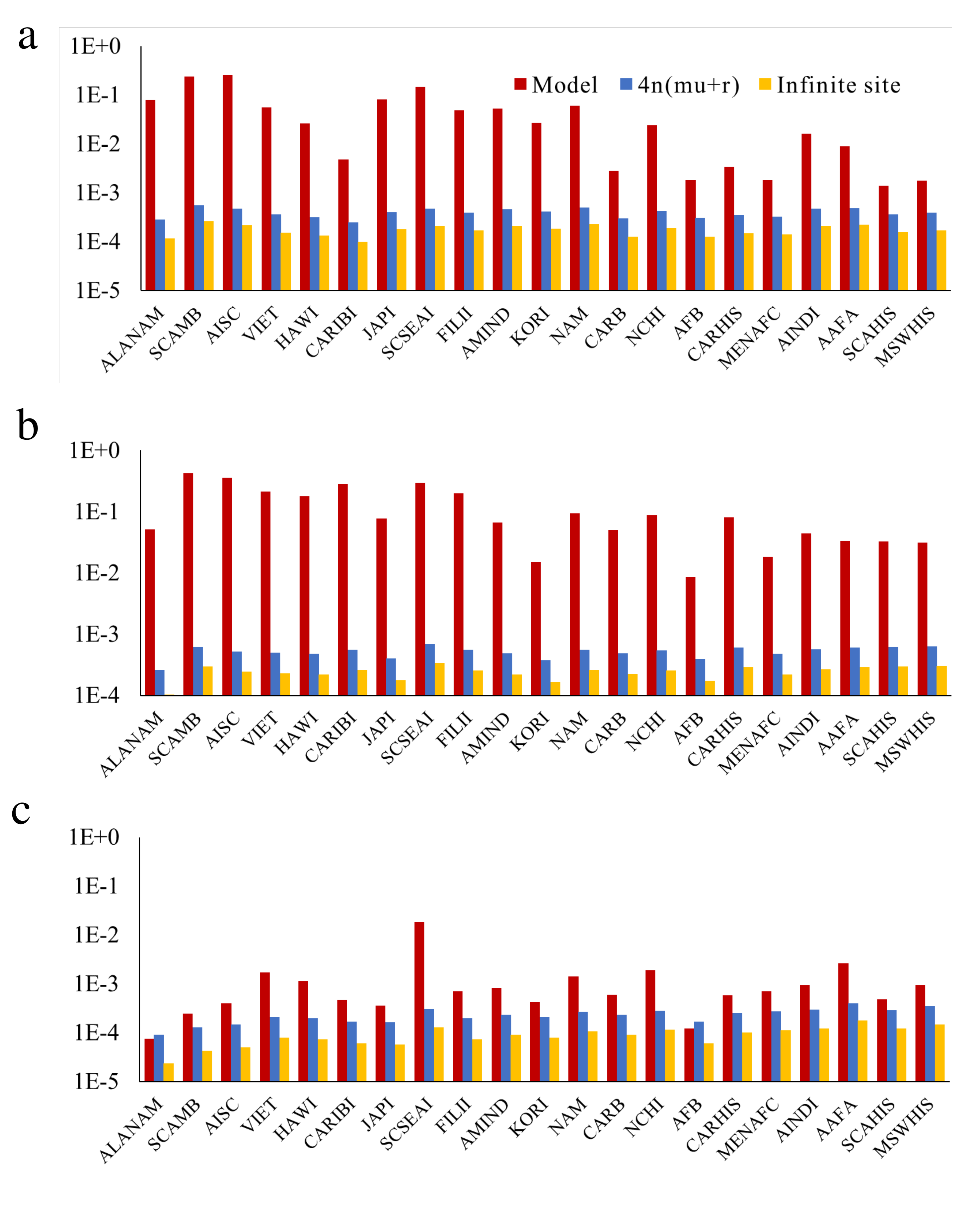}
\caption{Recombination rates across populations in the HLA region computed with our estimate, the infinite site model, and Watterson's estimator, for the pair of adjacent genes C-B (a), B-DR (b), and DR-DQ (c).}
\label{fig:populations}
\end{center}
\end{figure}

\bibliographystyle{apsrev4-1}

\end{document}